\documentclass[aps,prl,twocolumn,showpacs,showkeys,footinbib,longbibliography,groupedaddress]{revtex4-1}

\usepackage{amsmath}
\usepackage{amssymb}
\usepackage{graphicx}
\usepackage{bm}
\usepackage{color}
\usepackage{relsize}
\usepackage{braket}
\usepackage{bbold}
\usepackage{txfonts}
\usepackage{epstopdf}
\usepackage{hyperref}

\begin{document}

\title{Phase Control of Majorana Bound States in a Topological X Junction}

\author{Tong Zhou$^1$}
\email{tzhou8@buffalo.edu}
\author{Matthieu C. Dartiailh$^{2}$}
\author{William Mayer$^{2}$}
\author{Jong E. Han$^1$}
\author{Alex Matos-Abiague$^{3}$}
\author{Javad Shabani$^{2}$}
\author{Igor \v{Z}uti\'c$^{1}$}
\email{zigor@buffalo.edu}
\address{$^1$Department of Physics, University at Buffalo, State University of New York, Buffalo, New York 14260, USA\\
$^2$Center for Quantum Phenomena, Department of Physics, New York University, New York 10003, USA\\ 
$^3$Department of Physics and Astronomy, Wayne State University, Detroit, Michigan 48201, USA}
\date{\today}
\begin{abstract}
Topological superconductivity supports exotic Majorana bound states (MBS) which are chargeless zero-energy emergent quasiparticles. With their non-Abelian exchange statistics and fractionalization of a single electron stored nonlocally as a spatially separated MBS, they are particularly suitable for implementing fault-tolerant topological quantum computing. While the main efforts to realize MBS have focused on one-dimensional systems, the onset of topological superconductivity requires delicate parameter tuning and geometric constraints pose significant challenges for their control and demonstration of non-Abelian statistics. To overcome these challenges, building on recent experimental advances in planar Josephson junctions (JJs), we propose a MBS platform of X-shaped JJs. This versatile implementation reveals how external flux control of the superconducting phase difference can 
generate and manipulate multiple MBS pairs to probe non-Abelian statistics. The underlying topological superconductivity exists over a large parameter space, consistent with materials used in our fabrication of such X junctions, as an important step towards scalable topological quantum computing.
\end{abstract}
\maketitle

Majorana bound state (MBS), which are their own antiparticles~\cite{Kitaev2001:PU} are usually sought in proximity-modified 
materials~\cite{Fu2008:PRL,Lutchyn2010:PRL,Oreg2010:PRL,Alicea2010:PRB,Klinovaja2012:PRL,Zutic2019:MT}.
This approach overcomes the need for native spinless $p$-wave superconductivity, whose existence is debated even among the leading 
candidates such as Sr$_2$RuO$_4$~\cite{Mackenzie2003:RMP,Zutic2005:PRL}. Early MBS proposals have focused on proximity effects in 
one-dimensional (1D) geometries~\cite{Mourik2012:S,Rokhinson2012:NP,Deng2012:NL,Lee2012:PRL}. 
However, they rely on signatures, such as a quantized zero-bias peak~\cite{Sengupta2001:PRB,Law2009:PRL}, which do not probe the non-Abelian statistics crucial for implementing topological quantum computing~\cite{Nayak2008:RMP,Aasen2016:PRX}. These 1D platforms 
also pose inherent difficulties for MBS braiding. 

\begin{figure}[h]
\vspace{-0.05cm}
\includegraphics*[width=0.43\textwidth]{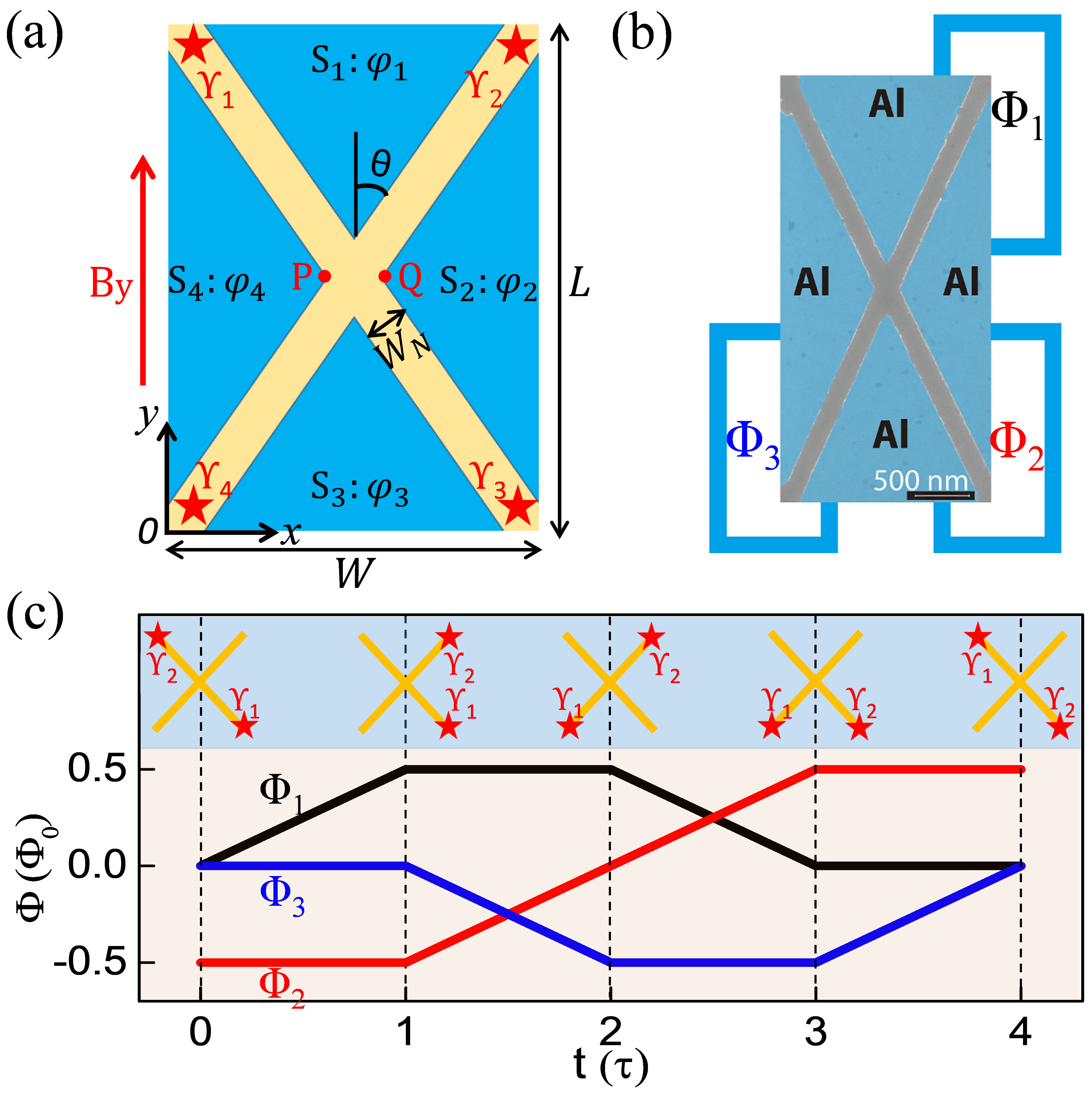}
\vspace{-0.34cm}
\caption{(a) X junction (XJ) schematic formed by  
epitaxial superconducting (S) regions (blue) covering a 2DEG (yellow).
A crossed X channel with the angle of $2\theta$, defined between the S leads, can be tuned into the topological regime with MBS 
$\gamma_1,...,\gamma_4$ (stars) at its ends by the phase differences among the S regions with an in-plane magnetic field 
along y-direction. All MBS pair combinations are 
obtained through modulating the phase differences between $\varphi_1$, $\varphi_2$, $\varphi_3$, and $\varphi_4$, 
supporting phase-controlled MBS manipulation. 
(b) SEM image for the XJ with schematic external fluxes $\Phi_1$ (between S$_1$, S$_2$), $\Phi_2$ (S$_2$, S$_3$), and $\Phi_3$ (S$_3$, S$_4$). 
(c) Exchange of MBS with external fluxes, $\tau$ is the switching time, and $\Phi_{0}$ is the magnetic flux quantum.}
\label{fig:XJF1}
\end{figure}
 
A push to seek alternative platforms for topological superconductivity led to the  demonstration of robust proximity-induced superconductivity 
in a two-dimensional electron gas (2DEG)~\cite{Shabani2016:PRB,Mayer2019:APL,Mayer2019:P1} which offers important opportunities to realize and control MBS~\cite{Pientka2017:PRX,Hell2017:PRL,Fatin2016:PRL,Stern2019:PRL,Scharf2019:PRB,Liu2019:P,Zhou2019:PRB,Setiawan2019:PRB,Setiawan2019:PRB2}. 
Recent experiments in planar Josephson junctions (JJs)~\cite{Fornieri2019:N,Ren2019:N,Mayer2019:P2}  reveal that topological superconductivity 
exists over a large parameter space without requiring fine tuning, while demonstrating the key role of superconducting phase to control the topological transition.
 
In this work we propose topological X-shaped junctions (XJs) as a natural 2D generalization of planar JJs, based on common normal (N) and superconducting (S) regions, to realize multiple MBS pairs and enable probing non-Abelian statistics. The setup is shown in Fig.~1(a). 
Four S films (Al) on the top of a 2DEG (InAs) form an X-shaped channel with the angle of 2$\theta$, defined between the S leads.  Such XJ,  
with an in-plane magnetic field, {\bf B}, along the y-direction, can be tuned into the topological regime by controlling the phase differences among 
the four S regions ($\varphi_1$,..., $\varphi_4$),  where either MBS pairs or  MBS quadruplet are formed at the XJ's ends. 

Experimentally, as seen from the scanning electron microscope image in Fig.~1(b), XJs can be fabricated and patterned easily using standard electron beam lithography by selectively removing Al. This technique was successful for fabrication of S/N and Josephson junctions~\cite{Kjaergaard2016:NC,Mayer2019:P2}. The top-down approach of XJs allows arbitrary design dimensions while keeping the channel between S electrodes less than the 2DEG's mean free path (within 100 nm for parameters considered). 
With the three flux loops [Fig.~1(b)], the phases between the four S regions can be tuned to enable moving and exchanging MBS. 
We show schematically exchanging MBS with the flux-control of $\Phi_1$, $\Phi_2$, and $\Phi_3$ in Fig.~1(c).

In the previous work on planar JJs, the focus was on {\bf B} applied along the S/N interface to realize 
MBS~\cite{Pientka2017:PRX,Fornieri2019:N}. However, 
to support multiple MBS in XJs, the topological superconductivity should survive when the in-plane {\bf B} deviates from that 
direction, just as B$_y$ in Fig.~1(a) is no longer along the S/N interface, forming the misalignment angle $=\theta$, discussed in 
Ref.~\cite{XJ_SM}.  
For a single JJ, our calculation shows that the phase-difference range supporting MBS become smaller when the misalignment angle
increased, but even  up to $0.15\pi$, there is still a finite phase-difference range (0.58$\pi$ to 1.16$\pi$) for topological states~\cite{XJ_SM}. 
As the misalignment angle increases, the topological region is reduced and, eventually, fully suppressed~\cite{XJ_SM}, as observed experimentally~\cite{Mayer2019:P2}.

The MBS robustness 
in a single JJ against {\bf B} misalignment angle $<0.2\pi$ provides guidance to design XJs~\cite{XJ_SM}.
We expect that XJs would support multiple MBS for $\theta < 0.2\pi$ with
the applied B$_y$, as shown in Fig.~1(a). The XJ can be viewed as including four planar JJs: S$_1$/N/S$_2$,  
S$_2$/N/S$_3$, S$_3$/N/S$_4$, and S$_4$ /N/S$_1$. 
For each JJ,  changing the phase difference from 0 to  $\phi_0$ ($\phi_0$ is studied in Fig.~2)  
results in a corresponding transition from trivial to topological junction with MBS at its ends.

At the XJ's center, common to individual JJs, MBS can be fused.  We distinguish their three types: short-edge ($\gamma_{3}$, $\gamma_4$) with the phases ($\varphi_1$, $\varphi_2$, $\varphi_{3}$, $\varphi_{4}$) = (0, 0, $\phi_{0}$, 0), 
long-edge ($\gamma_2$, $\gamma_3$) with (0, $\phi_0$, 0, 0), and diagonal  ($\gamma_1$, $\gamma_3$) with 
($\phi_0$, $\phi_0$, 0, 0), as shown in Fig.~1(a). To study MBS control and identify the $\phi_0$ supporting topological states, in our calculations we use the Bogoliubov-de Gennes (BdG) Hamiltonian, 
\begin{equation}
\begin{aligned} 
H = \left[\frac{\mathbf{p}^2}{2m^\ast} - \mu\left(x,y\right) + \frac{\alpha}{\hbar}\left(p_y\sigma_x - p_x\sigma_y \right)\right]\tau_z\\ - \frac{g^\ast\mu_B}{2}\mathbf{B}\cdot\boldsymbol{\sigma} + \Delta\left(x,y\right)\tau_+ + \Delta^\ast\left(x,y\right)\tau_-\;,
\end{aligned} 
\label{eq:BdG}
\end{equation}
numerically solving the corresponding eigenvalue problem 
on a discretized lattice as implemented in Kwant~\cite{Groth2014:NJP}.
Here $\mathbf{p}$ is the momentum, $\mu(x, y)$ is the chemical potential, $\alpha$ is the Rashba SOC strength, $\mathbf{B}$ is the external magnetic field, $\mu_B$ is the Bohr magneton,  while $m^\ast$  and $g^\ast$ are the electron effective mass and $g$-factor, respectively. 
We use $\tau_i$ $(\sigma_i)$ as the Nambu (Pauli) matrices in particle-hole (spin) space and $\tau_\pm  = (\tau_x \pm \tau_y)/2$. 
$\Delta(x, y)$ is the proximity-induced superconducting pair potential, for the 2DEG below the S leads S$_n$ 
(n = 1,...,4), it can be expressed, using the BCS relation for the B-field suppression, as
\begin{equation}
\Delta(x, y) = \Delta_{0} \sqrt{1-\left(B / B_{c}\right)^{2}}e^{i\varphi_n}, 
\label{eq:delta}
\end{equation} 
where  $\Delta_0$ is the superconducting gap, B$_c$ is the critical magnetic field, and  $\varphi_n$ is the corresponding superconducting phase.
The S and N regions are simply expressed using the coordinates of the points P and Q in Fig.~1(a). For example,    
\begin{equation}
S_{1}(x, y)=\left\{\begin{array}{c}{y>\cot \theta\left(x-x_{P}\right)+y_{P}} \\ {y>-\cot \theta\left(x-x_{Q}\right)+y_{Q}}\end{array}\right.,
\end{equation} 
where $x_{P,Q} = (W \mp W_N/\cos \theta)/2$, 
$y_{P,Q} = L/2$.  The other S and N regions are explicitly given in Ref.~\cite{XJ_SM}.
We choose materials parameters consistent with our fabricated XJs [Fig.~1(b)] that also 
match experimental observation of robust proximity-induced superconductivity and topological states in  epitaxial Al/InAs-based JJs~\cite{Mayer2019:P2}, $m^\ast=0.03 m_0$, where $m_0$ is the electron mass, and $g^\ast=10$
for InAs, $\Delta_0$ = 0.23 meV, $\alpha$ = 10 meVnm, B$_c$ = 1.6 T, and for the N, S chemical potential   
$\mu_{S}$ = $\mu_{N}$ = 0.5 meV. 
We consider XJ with  $L=3.2$ $\mu$m, $W=L/2$, $W_N\approx100$ nm, and Al/InAs materials parameters in Figs.~2-4.

\begin{figure}
\centering
\includegraphics*[width=0.45\textwidth]{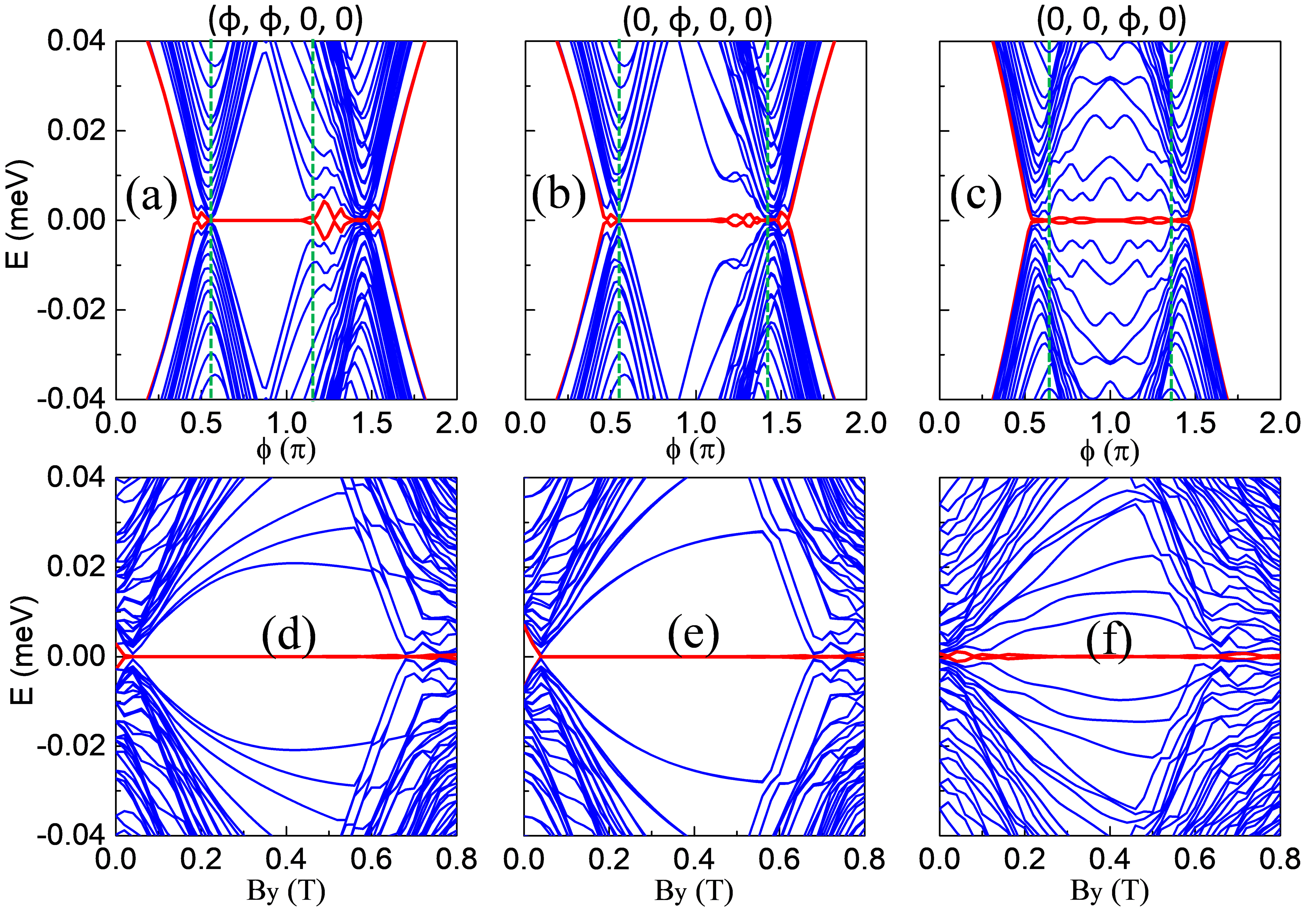}
\caption{(a)-(c) Energy spectra at B$_y=0.4$ T as a function of $\phi$ for the diagonal, long- and short-edge MBS in the XJ with 
half-central angle $\theta$ = 0.1$\pi$, as seen in Fig.~1.
The red lines: evolution of finite-energy states into MBS inside the topological gap. The $\phi$ values between the two green lines ($\phi_0$) give the topologically nontrivial states. (d)-(f) Magnetic field dependence of the energy spectrum with superconducting phases ($\varphi_1,...,\varphi_4$) = ($\pi$, $\pi$, 0, 0), (0, $\pi$, 0, 0), and (0, 0, $\pi$, 0), respectively. The red lines retain the meaning as in (a)-(c). The materials and geometry parameters are specified in the main text.}
\label{fig:XJF2}
\end{figure}

It is instructive to examine the robustness of the topological states  as a function of $\phi$ for the diagonal, long- and short-edge MBS, 
shown in Figs.~2(a)-(c), respectively. The evolution of the lowest-energy states into MBS
reveals a large 
range of $\phi_{0} \in (0.6\pi, 1.2\pi)$ for diagonal, $(0.6\pi, 1.4\pi)$ for long-edge, and $(0.64\pi,1.36\pi)$ for short-edge MBS. 
Since the geometry for diagonal MBS
($\varphi_1$ = $\varphi_2$ = $\phi$, and $\varphi_3$ = $\varphi_4$ = 0) resembles the single JJ with the B-field misalignment $\theta$,
we also expect the similarities in their spectra. Indeed, for $\theta$ = 0.1$\pi$ this
can seen in Ref.~\cite{XJ_SM}.

The results from Figs.~2(a)-(c) suggest that, as in a single JJ~\cite{Pientka2017:PRX}, the choice $\phi_0 = \pi$ is particularly desirable 
for the stability of the topological state and could reduce the critical field for the onset of MBS in XJs. For such $\phi_0 = \pi$  we also examine
complementary information about the robustness of topological states with in-plane B$_y$. From the low-energy spectra in Figs.~2(d)-(f),
we see that, similar to the single JJ, a small B$_y \sim 0.1$ T  
already 
supports all three MBS types. As expected from MBS in short 1D 
systems, for the short-edge [Fig.~2(f)] the topological gap is the smallest and its zero energy-bands have small oscillations which can be
suppressed with an increased system size~\cite{XJ_SM,Lim2012:PRB,Prada2012:PRB,DasSarma2012:PRB,Rainis2013:PRB}.

\begin{figure*}[t]
\centering
\includegraphics*[width=0.75\textwidth]{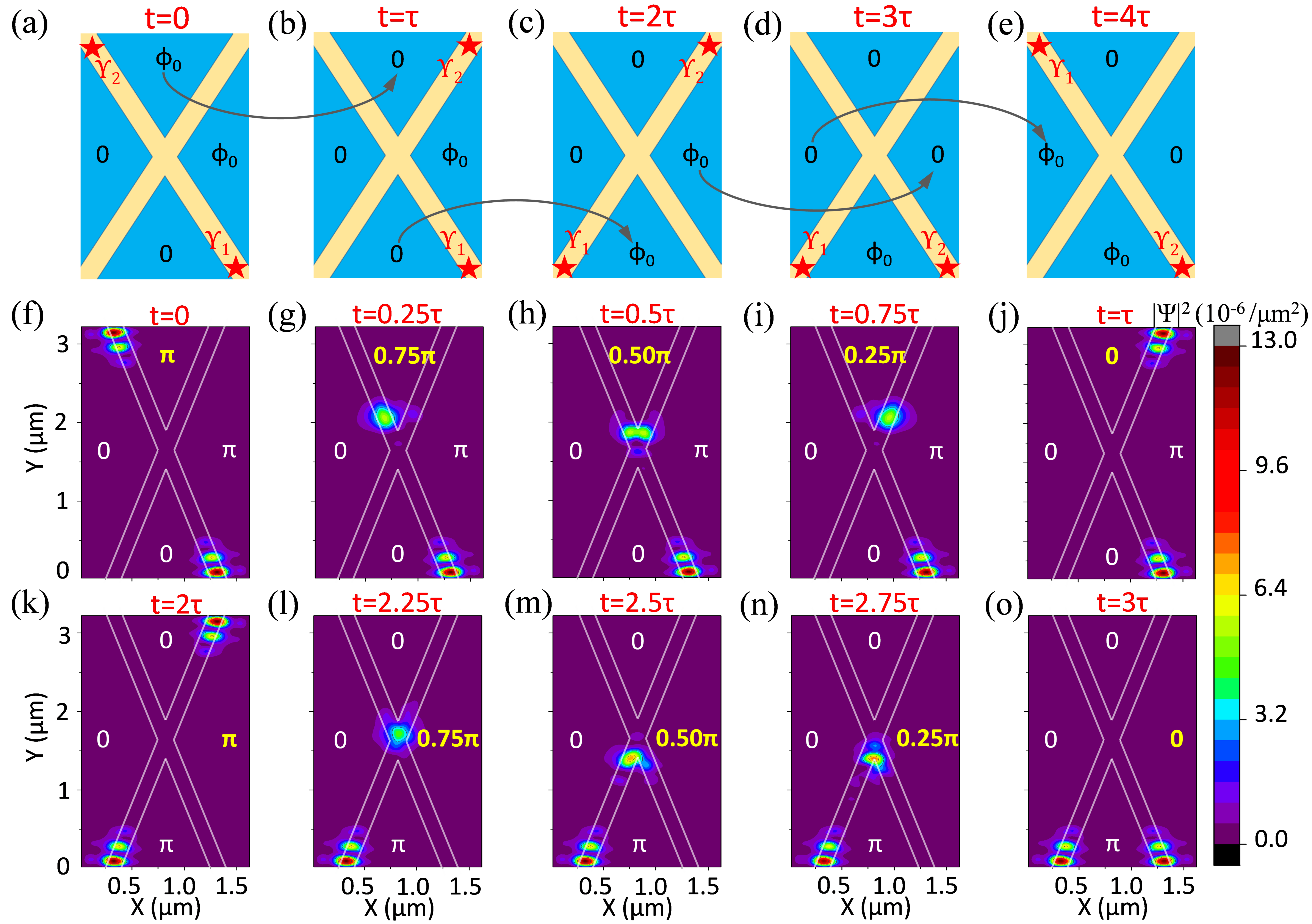}
\caption{(a)-(e) Schematic of exchanging MBS in the XJ with external flux control, following the protocol from Fig.~1(c).
$\phi_{0}$ is the phase difference supporting all three MBS types shown in Fig.~2. (f)-(j) Evolution of the calculated MBS probability densities from the diagonal (a) to a long-edge MBS (b) through continuously changing $\varphi_1$ from $\pi$ to 0, but fixing $\varphi_2$ = $\pi$, $\varphi_3$ = 0, and $\varphi_4$ = 0. (k)-(o) Evolution of the calculated MBS probability densities from diagonal (c) to short-edge MBS (d) by changing $\varphi_2$ from $\pi$ to 0, but fixing $\varphi_1$ = 0, $\varphi_3$ = $\pi$, and  $\varphi_4$ = 0. It is not important if the corners do not coincide with the ends of diagonals. 
The Al/InAs parameters are taken from Fig.~2 with B$_y = 0.2$ T.} 
\label{fig:XJF3}
\end{figure*}

While our previous results are encouraging, suggesting that different MBS could be supported in XJs, they all considered a fixed half-angle between 
the two N regions, $\theta=0.1\pi$ which could be challenging to exactly replicate experimentally. It is therefore important to indentify the 
$\theta$-range still allowing these topological states. With the fixed system size ($L=3.2$ $\mu$m, $W=L/2$, and $W_N \approx 100$ nm) in 
Ref.~\cite{XJ_SM} we calculate $\theta$-dependent energy spectra for different MBS. 
When $\theta\sim 0$ the diagonal and long-edge configuration can be approximated as a single $\pi$-JJ where the MBS are stable
up to a misalignment angle $\sim 0.2\pi$ of the B-field away from the N/S interface. 
Indeed, when related XJs are calculated we find that these two MBS types are supported for $\theta \lesssim 0.18\pi$.
However, for short-edge MBS and $\theta\sim 0$, the areas of S$_1$ and S$_3$ are too small to support such MBS. Instead, MBS 
emerge at $\theta \approx 0.08\pi$ and remains stable for $\theta \lesssim 0.18\pi$~\cite{XJ_SM}. 

With these results, we can identify that all the three MBS types can coexist in a robust form for $\theta \in (0.08\pi, 0.18\pi)$ with $L$ $\geq$ 3.2 $\mu$m and $W$ $\geq$ 1.6 $\mu$m. Such a large parameter range gives a considerable flexibility for XJ fabrication.  
Our 
fabricated XJ [Fig.~1(b)] 
already fits well in this range of suitable geometric 
parameters, with $L$ = 4.0 $\mu$m, $W$ = 2.0 $\mu$m and $\theta$ = 0.15$\pi$.

The existence of MBS is an important precondition, but alone does not ensure their successful manipulation in XJs which could 
offer an important generalization of the previous planar JJs to a much more versatile 2D platform. We therefore consider in detail the scheme for 
phase control of MBS with three external fluxes sketched in Fig.~1(c) such that flux in one of the four S regions is changed
within the switching time, 
$0.2 \: \mathrm{ns} < \tau \sim 1\mu$s. The lower and upper limits are given by the uncertainty relation using the smallest calculated energy 
gap $\sim 2$ $\mu$eV to the first excited state~\cite{XJ_SM} and the quasiparticle poisoning time~\cite{Albrecht2017:PRL,Rainis2012:PRB,Aseev2018:PRB}, respectively.

Since two MBS can exist at any two ends of the XJ with the specific phase differences, it gives us a chance to realize MBS exchange and fusion using the phase control. To describe the MBS exchange of $\gamma_1$ and $\gamma_2$ in XJs, we plot schematically their evolution in Figs.~3(a)-(e). 

Initially, at t = 0, diagonal MBS ($\gamma_1$, $\gamma_2$) 
are located at the opposite corners, depicted in Fig.~3(a), by setting the three fluxes ($\Phi_1$, $\Phi_2$, $\Phi_3$) as (0, -0.5$\Phi_0$, 0). 
First, $\gamma_2$ is moved to the upper-right corner to form the long-edge MBS shown in Fig.~3(b) by 
changing $\Phi_1$ from 0 to 0.5$\Phi_0$ at t = $\tau$. The feasibility of this process is verified by calculating the adiabatic evolution
of the MBS probability densities, shown in Figs.~3(f)-(j), for the realistic Al/InAs parameters from Fig.~2 at  B$_y =0.2$ T. 
During the switching time $\tau$ , when the phase difference $\varphi_1$ changes continuously from 0 to $\phi_0=\pi$, 
well-localized MBS are protected by the topological superconducting gap such that $\gamma_2$ can be adiabatically 
moved from the upper-left to the upper-right corner through the XJ center~\cite{XJ_SM}.

In the next step, $\gamma_1$ is moved from the lower-right to the lower-left corner in Fig.~3(c) by changing ($\Phi_2$, $\Phi_3$) 
from  (-0.5$\Phi_0$, 0) to (0, -0.5$\Phi_0$) at t = 2$\tau$. This process changes the long-edge to a diagonal MBS, 
which can, because of the mirror symmetry in XJ, be viewed as equivalent to the transition from Fig.~3(b) to 3(a). The following step is to move $\gamma_2$ from the upper-right to the lower-right corner in Fig. 3(d) by changing ($\Phi_1$, $\Phi_2$) from (0.5$\Phi_0$, 0) to (0, 0.5$\Phi_0$) at t = 3$\tau$.  
Changing the diagonal to a short-edge MBS
is verified by calculating the MBS probability densities, shown in Figs.~3(k)-(o), where the phase difference $\varphi_2$ changes from  $\pi$ to 0.
We can see the MBS are robust and $\gamma_{2}$ can be adiabatically moved to the original position of $\gamma_1$. Finally, $\gamma_1$ is moved from the lower-left to the upper-left corner in Fig.~3(e) by changing $\Phi_3$ from -0.5$\Phi_0$ to 0 at t = 4$\tau$. 
This process, due to the mirror symmetry in XJ, is equivalent  to the transition from Fig.~3(d) to (c), where $\gamma_1$ is moved to the original position of 
$\gamma_2$. Through these four switching steps (from t = 0 to t = 4$\tau$), $\gamma_1$ and $\gamma_2$ are adiabatically exchanged.
Further information about this exchange, including related movies, is in Ref.~\cite{XJ_SM}.

\begin{figure}[t]
\centering
\includegraphics[width=0.46\textwidth]{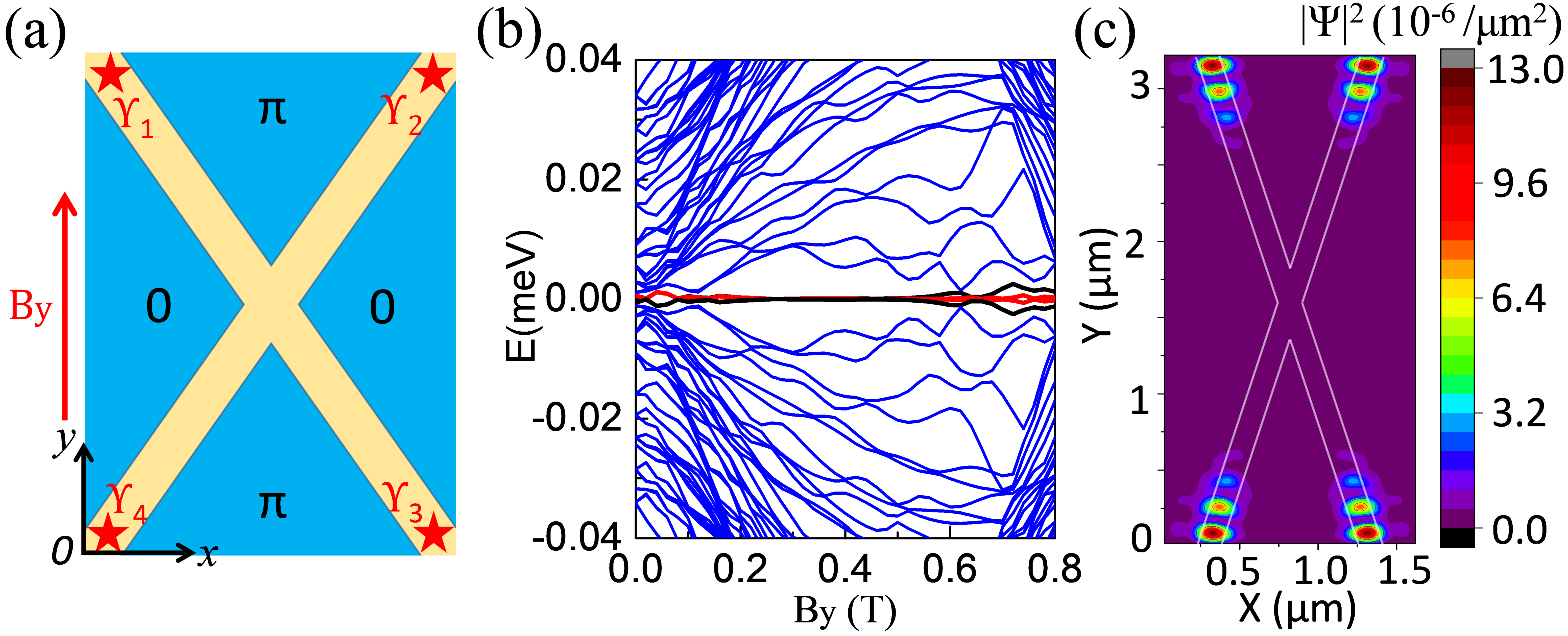}
\caption{(a) Schematic of four simultaneous MBS in the XJ with superconducting phases  ($\varphi_1$, $\varphi_2$, $\varphi_3$, $\varphi_4$) = ($\pi$, 0, $\pi$, 0). (b) Magnetic field dependence of the low-energy spectrum for (a). (c) Calculated probability densities for the two lowest-energy states in (b) with 
B$_y=0.2$ T. The parameters are taken from Fig.~2.}
\label{fig:XJF4}
\end{figure}

Our proposed XJs can be used to explore the MBS fusion rules~\cite{Aasen2016:PRX}. 
For example, starting from the diagonal MBS in Fig. 3(a), $\gamma_1$ and $\gamma_2$ can be gradually moved and fused at the center of 
the XJ by using the three external fluxes to change $\varphi_1$ from 0 to 0.5$\pi$, and then to change $\varphi_3$ from $\pi$ to 0.5$\pi$, within the time $2\tau$. The corresponding evolution of the calculated MBS densities is shown in Ref.~\cite{XJ_SM}.
Since we have shown in Fig.~4 that 4 MBS can be generated, a more complex fusion in XJ, for example, fusing ($\gamma_1$, $\gamma_2$) and fusing ($\gamma_2$, $\gamma_3$) can access different fusion channels, which could probe non-Abelian statistics~\cite{Aasen2016:PRX}.

The versatility of the MBS phase-control in 2D XJs also provides a powerful platform to implement braiding which requires at least four MBS. 
In a single XJ, as shown in Fig.~4(a), four  MBS can be simultaneously realized at the four corners when $(\varphi_1, \varphi_2, \varphi_3, \varphi_4) 
= (0, \pi, 0, \pi)$ set by the fluxes $(\Phi_1, \Phi_2, \Phi_3) = (0.5\Phi_0, 0.5\Phi_0, -0.5\Phi_0)$. The presence of these MBS is verified from the calculated low-energy spectra in Fig. 4(b). The two MBS pairs exist within the topological gap for B$_y \in  (0.1$ T,  0.6 T$)$. 
The corresponding wavefunction probability densities also clearly indicate in Fig.~4(c) that the formation of four MBS are localized at the four corners of the XJ. An important advantage of our platform is its scalability, using lithography, multiple XJs can be realized.
In just two joint XJs having six ends in the N regions at which MBS can be localized, with external fluxes and phase control fifteen different realizations of four MBS are possible. 

While our work was motivated by the recent advanced in high-quality epitaxial Al/InAs junctions several of our findings have direct implications for other
systems seeking to manipulate Majorana bound states. For example, state-of-the art fabrication of superconducting junctions with topological insulators~\cite{Schuffelgen2019:NN} would support fabrication of similar X-shaped junctions. With the progress in tunable magnetic textures used
to modify proximity-induced superconductivity~\cite{Fatin2016:PRL,Matos-Abiague2017:SSC,Zhou2019:PRB,Wei2019:PRL,Kjaergaard2012:PRB,Mohanta2019:PRA,
Desjardins2019:NM,Yazdani2019:NM,Virtanen2018:PRB,Kim2015:PRB,Marra2017:PRB,Gungordu2018:PRB,Yang2016:PRB,Palacio-Morales2019:SA}, 
X-shaped junctions could be considered with a reduced role of an applied magnetic field.

Throughout our calculations the influence of the central angle $2 \theta$ between two S/N interfaces was emphasized, which could be further optimized~\cite{Boutin2018:PRB}. While common approaches to realize exchanging and braiding envision structures such as the 
T-junction or crossbar geometries~\cite{Alicea2011:NP}, we see their underlying right angles as detrimental to the robust manipulation 
of Majorana bound states. Instead, as in  our X-shaped junctions, it is important to implement acute angles $2\theta < \pi/2$ in schemes relying on 
applied in-plane magnetic field. In an early work on multiple Josephson junctions with topological insulators~\cite{Fu2008:PRL} the situation is slightly better with the characteristics $\pi/3$ angle. However, that angle is still too large for junctions with more common materials, such as Al/InAs,  using in-plane magnetic field. 
Experimental demonstration of the proposed two-dimensional manipulation of Majorana states in these topological
X-shaped junctions would constitute an important milestone towards scalable topological quantum computing and stimulate further studies 
in the design of emergent phenomena in proximitized materials~\cite{Klinovaja2014:PRB}. 
 
\begin{acknowledgements}
This work is supported by DARPA Grant No. DP18AP900007, US ONR Grant No. N000141712793 (I. \v{Z}. and A. M.-A.), 
and the UB Center for Computational Research.
\end{acknowledgements} 

\bibliography{XJ_bib}

\end{document}